\documentclass[prb,superscriptaddress,twocolumn,preprintnumbers,amsmath,floatfix,showpacs]{revtex4}
\usepackage{amssymb}
\usepackage{graphicx}

\begin{document}

\title{Hidden $(\pi,0)$ instability as an itinerant origin of bicollinear antiferromagnetism in Fe$_{1+x}$Te}

\author{Ming-Cui Ding}
\affiliation{Shanghai Key Laboratory of Special Artificial Microstructure Materials and Technology, \\
$\&$ School of Physics Science and engineering, Tongji University, Shanghai 200092, P.R. China}
\author{Hai-Qing Lin}
\affiliation{Beijing Computational Science Research Center, Beijing 100084, China}
\author{Yu-Zhong Zhang}
\email[Corresponding author. ]{Email: yzzhang@tongji.edu.cn}
\affiliation{Shanghai Key Laboratory of Special Artificial Microstructure Materials and Technology, \\
$\&$ School of Physics Science and engineering, Tongji University, Shanghai 200092, P.R. China}

\date{\today}

\begin{abstract}
By calculating orbitally resolved Pauli susceptibilities within maximally localized Wannier orbital basis transformed from first principles band structures, we find that magnetism in Fe$_{1+x}$Te still has its itinerant origin even without Fermi surface nesting, provided orbital modulation of particle-hole excitations are considered. This leads to strong magnetic instabilities at wave vector (0,$\pi$)/($\pi$,0) in d$_{xz}$/d$_{yz}$ orbitals that are responsible for the bicollinear antiferromagnetic state as extra electrons donated from excess Fe are considered. Magnetic exchange coupling between excess Fe and in-plane Fe further stabilizes the bicollinear antiferromagnetic order. Our results reveal that magnetism and superconductivity in iron chalcogenides may have different orbital origin, as Pauli susceptibilities of different orbitals evolve differently as a function of concentration of excess Fe and height of the chalcogen atom measured from the iron plane.
\end{abstract}

\pacs{74.70.Xa,71.20.-b,75.25.-j,74.25.Jb}

\maketitle

\section{Introduction}
Unconventional superconductivity emerges from suppression of the magnetically ordered state in most iron-based superconductors~\cite{Liu2010,Zhao2008,Kimber2009}. Thus, unveiling the origin of magnetism is a crucial step towards understanding the mechanism of superconductivity in these compounds~\cite{Paglione2010,Mazin2010}. In contrast to iron pnictides where magnetism can be explained from both itinerant~\cite{Mazin2008} and localized~\cite{Yildirim2008,Si2008} limit, magnetism in iron chalcogenides can only be accounted for by a localized scenario~\cite{Ma2009,Moon2010}, due to the fact that no Fermi surface nesting at the wave vector of $(\pi,0,\pi)$ is detected from angle resolved photoemission (ARPES) study~\cite{Xia2009,Nakayama2010}, which conflicts with the observation of bicollinear antiferromagnetic (BAF) order from neutron diffraction analysis~\cite{Bao2009,Li2009}. This remarkable difference between iron pnictides and iron chalcogenides is a serious challenge against establishing a unified theory for different families of iron-based superconductors~\cite{Kamihara2008,Rotter2008,Hsu2008}.

Han and Savrasov have attempted to recover the itinerant scenario for magnetism in Fe$_{1+x}$Te by assuming that each excess Fe contributes eight extra electrons into the Fe plane, leading to a substantial shifting of Fermi level and a dramatic change of Fermi surface topology~\cite{Han2009,Singh2010,Han2010}. A strong peak for condensation of particle-hole excitations emerges in the calculated Pauli susceptibility right at wave vector $(\pi,0,\pi)$ which accounts for the itinerant nature of BAF order. However, this doping effect was not observed in ARPES experiments~\cite{Xia2009,Nakayama2010} and the oxidation state of interstitial Fe should be close to that of in-plane Fe as implied by neutron diffraction~\cite{Rodriguez2011}. Furthermore, our density functional theory (DFT) calculations based on nonmagnetic and BAF states both support the fact that all the irons share a similar valence state (see appendix \ref{sec:one}). Therefore the shift of Fermi level due to the extra electrons donated from excess Fe is one order of magnitude smaller than that done by Han and Savrasov~\cite{Han2009,Singh2010,Han2010} (see appendix \ref{sec:two}). As a consequence, Fermi surface topology remains almost unchanged compared to the undoped case, leading to a good agreement with ARPES observations. However, the calculated Pauli susceptibilities within a constant matrix element approximation as usually done in the literature~\cite{Mazin2008,Dong2008,Heil2012} only show a notable peak at $(\pi,\pi,\pi)$ in both undoped and doped cases, which again casts doubts on the itinerant picture of magnetism in Fe$_{1+x}$Te (see appendices \ref{sec:three} and \ref{sec:four})

Here we will demonstrate that the ignored matrix elements in calculated Pauli susceptibility are the key quantities for understanding the origin of itinerant magnetism in Fe$_{1+x}$Te. The particle-hole excitation is strongly modulated by orbitals, and its condensation wave vector is orbitally dependent. Within a reasonable amount of extra electrons donated from excess Fe, prominent magnetic instability at (0,$\pi$)/($\pi$,0) in d$_{xz}$/d$_{yz}$ orbital is obtained. The phase transition from commensurate to incommensurate antiferromagnetic state observed experimentally can also be naturally explained as excess Fe further increases. Our results reveal that multiple instabilities coexist in different orbitals and evolve differently as a function of extra electrons and height of chalcogen atom measured from the iron plane, suggesting that magnetism and superconductivity in iron chalcogenides may have different orbital origin.

\begin{figure*}[tb]
\includegraphics[width=\textwidth]{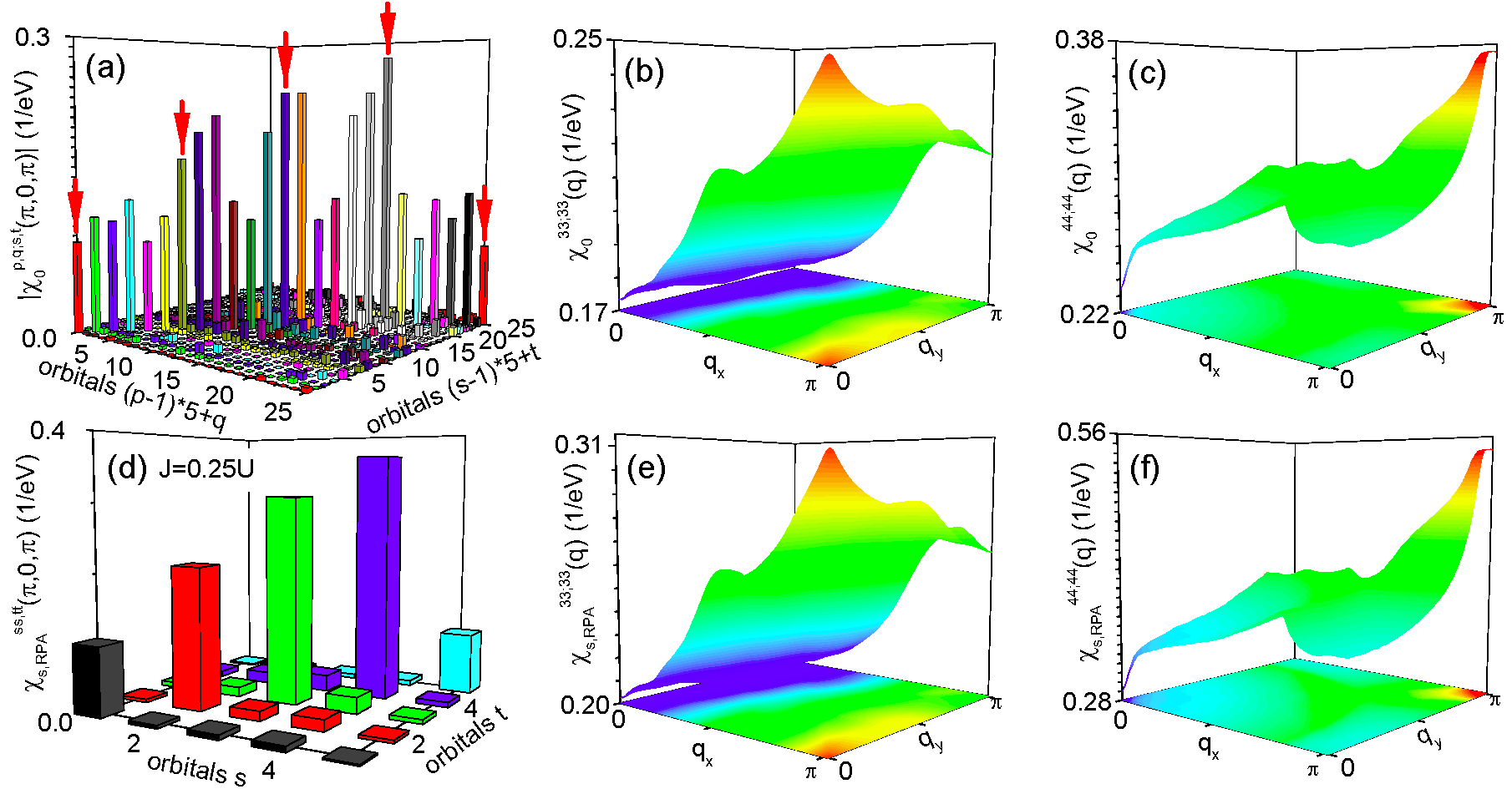}
\caption{(Color online) Orbitally resolved Pauli susceptibility and spin susceptibility. Magnitudes of all the elements in Pauli susceptibility (\textbf{a}) and spin susceptibility (\textbf{d}) at $q=(\pi,0,\pi)$. Dominant q-dependent Pauli susceptibility $\chi^{33;33}_{0}(q,\omega=0)$ (\textbf{b}), $\chi^{44;44}_{0}(q,\omega=0)$ (\textbf{c}) at $q_z=\pi$. Dominant q-dependent spin susceptibility $\chi^{33;33}_{s,RPA}(q,\omega=0)$ (\textbf{e}), $\chi^{44;44}_{s,RPA}(q,\omega=0)$ (\textbf{f}) at $q_z=\pi$. The two-dimensional contour maps are on the bottom. Here Fermi level is shifted up by 0.075eV, corresponding to the Fe$_{1+x}$Te compound at $x=0.10$~\cite{Figureref}. Spin susceptibility is calculated at $J=U/4$ and $U=0.8$ eV. The diagonal elements of Pauli susceptibility relevant to magnetism are marked by arrows in (\textbf{a}). From left to right, those are $\chi^{tt;tt}_{0}(q,\omega=0)$ with $t$ from $1$ to $5$.} \label{fig:Pauli3D}
\end{figure*}

\section{details of our calculations}
We use the experimental lattice structure of the paramagnetic phase~\cite{Li2009} throughout the paper, except when we study the substitution effect of Te by Se, which can be effectively viewed as a reduction of the height of the chalcogen atom measured from the Fe plane. We employ the full potential linearized augmented plane wave method as implemented in WIEN2K~\cite{Blaha2001} to calculate the electronic properties of Fe$_{1+x}$Te. An effective d-p model of 16 orbitals within maximally localized Wannier basis~\cite{Wannier90,Wientowannier} is constructed in order to calculate orbitally resolved Pauli susceptibility and spin susceptibility. We use a three-dimensional grid of 128$\times$128$\times$128 k and q points in the whole Brillouin zone with a temperature smearing of 0.01 eV. Long range hoppings are involved so that both band structure and density of states from effective model are perfectly consistent with those from first principles in the energy window of [-6 eV, 3 eV] (see appendices \ref{sec:seven} and \ref{sec:eight})

The exact Pauli susceptibility~\cite{Graser2009} at the Fermi level $\chi^{pq;st}_{0}(q,\omega=0)$ (defined in appendix \ref{sec:five}) is a four-index tensor with each index varying from 1 to 5 which represents five 3d orbitals of Fe (orbital 1: d$_{xy}$, 2: d$_{xz}$, 3: d$_{yz}$, 4: d$_{x^2-y^2}$ , 5: d$_{z^2}$, where x, y, z refer to those for the original unit cell). In Fig.~\ref{fig:Pauli3D}~({\bf a}), a total of 625 elements of the Pauli susceptibility at the Fermi level are presented at $q=(\pi,0,\pi)$ in a doped case where the Fermi level is moved up by 0.075 eV. It is found that off-diagonal elements of Pauli susceptibility are negligibly small, compared to the diagonal ($p=s,q=t$) elements which represent intraorbital and interorbital particle-hole excitations. Moreover, the spin susceptibility, which determines magnetism, is only related to the elements of Pauli susceptibility with the indices of $s=t$. Due to the above two facts, only intraorbital particle-hole excitations, i.e., $\chi^{pq;st}_{0}(q,\omega=0)$ with $p=s=q=t$, as marked by arrows in Fig.~\ref{fig:Pauli3D}~({\bf a}), have to be taken into account in order to discuss possible magnetism, while other diagonal elements are irrelevant to the magnetism. We have also analyzed Pauli susceptibility at different wave vectors and in different doping cases including the undoped case and found that the above conclusion is valid in all cases.

\section{Results and discussions}
In Figs.~\ref{fig:Pauli3D}~({\bf b}) and~({\bf c}), two dominant elements of Pauli susceptibility $\chi^{tt;tt}_{0}(q,\omega=0)$ with $t=3,4$ are shown. The Fermi level is again shifted up by 0.075 eV in order to account for the excess electrons contributed from interstitial Fe. While a pronounced peak remains close to the wave vector $q=(\pi,\pi,\pi)$ in the d$_{x^2-y^2}$ orbital, a well-defined strong peak appears at $q=(\pi,0,\pi)$ in d$_{xz}$ orbitals, which is responsible for the BAF order. $\chi^{22;22}_{0}(q,\omega=0)$ (see appendix \ref{sec:fourx}) shows a mirror symmetry to $\chi^{33;33}_{0}(q,\omega=0)$ with respect to $q_x=q_y$, which exhibits a strong peak at $q=(0,\pi,\pi)$. $(\pi,0,\pi)$ and $(0,\pi,\pi)$ instabilities are also found in the d$_{xy}$ orbital, which is considerably smaller than those in the d$_{xz}$ and d$_{yz}$ orbitals. Similarly, the instability at $(\pi,\pi,\pi)$ in the d$_{z^2}$ orbital is much weaker than that in the d$_{x^2-y^2}$ orbital. These may indicate that d$_{xy}$ and d$_{z^2}$ orbitals play minor roles in the formation of magnetism or superconductivity.

\begin{figure}
\includegraphics[width=\columnwidth]{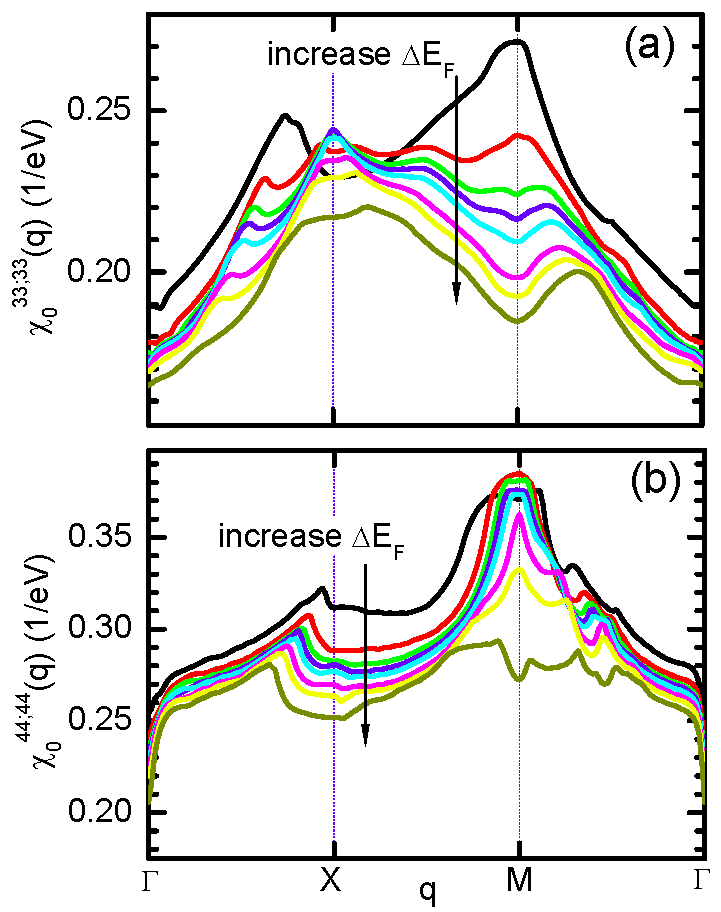}
\caption{(Color online) Evolution of particle-hole excitations in Pauli susceptibility at different shifted Fermi energies.(\textbf{a}) $\chi^{33;33}_{0}(q,\omega=0)$. (\textbf{b}) $\chi^{44;44}_{0}(q,\omega=0)$. $\Delta E_F=0,0.045,0.065,0.075,0.085,0.105,0.12,0.15$ eV, corresponding to the Fe$_{1+x}$Te compounds with $x=0, 0.06, 0.09, 0.10, 0.12, 0.14, 0.16, 0.20$~\cite{Figureref}.} \label{fig:ShiftEf}
\end{figure}

\begin{figure}
\includegraphics[width=\columnwidth]{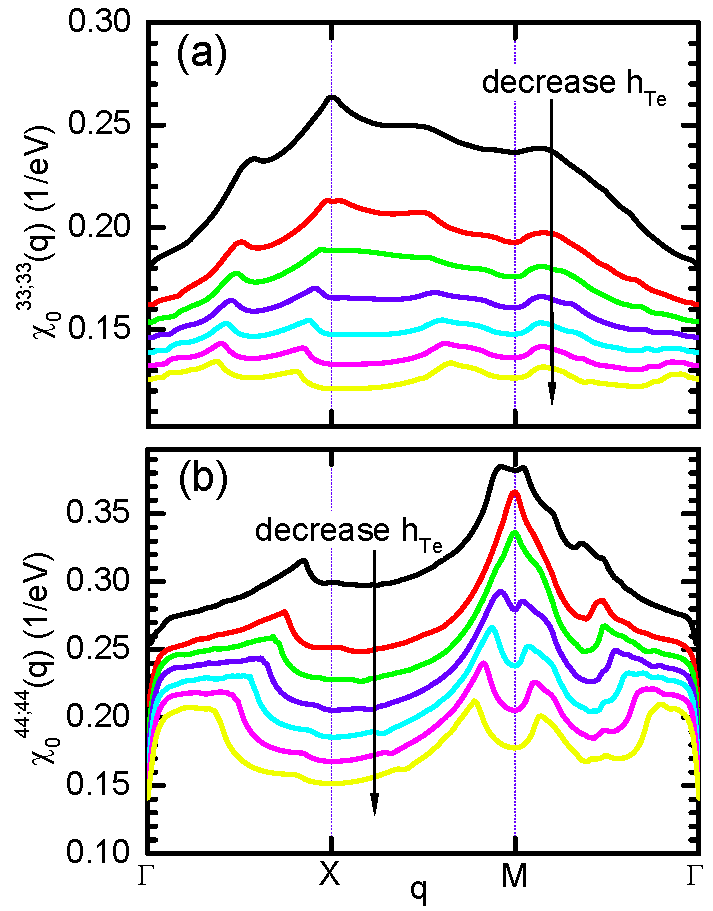}
\caption{(Color online) Evolution of particle-hole excitations in Pauli susceptibility as a function of Te-height measured from the Fe plane. (\textbf{a}) $\chi^{33;33}_{0}(q,\omega=0)$. (\textbf{b}) $\chi^{44;44}_{0}(q,\omega=0)$. Te height from $h_{Te}=1.8\AA$ to $1.52\AA$ in the interval of $0.04\AA$.} \label{fig:ShifthTe}
\end{figure}

Although it is interesting to get the fundamental $(\pi,0,\pi)$ and $(0,\pi,\pi)$ instability by a reasonable rigid band shift served as doping extra electrons to the Fe plane from interstitial Fe, one may still doubt whether this instability vanishes as local Coulomb interactions and Hund's rule couplings turn on. In Figs.~\ref{fig:Pauli3D}~({\bf d}),~({\bf e}) and~({\bf f}), we show the effect of interactions on the orbitally dependent instabilities. The spin susceptibility is a 5$\times$5 matrix and can be obtained in the form of Dyson-type equation within random phase approximation(RPA). (see appendix \ref{sec:five}) Fig.~\ref{fig:Pauli3D}~({\bf d}) presents all 25 elements of spin susceptibility at $J/U=0.25$ with $U=0.8eV$ and at $q=(\pi,0,\pi)$. It is found that off-diagonal elements are negligibly small, compared to the diagonal ones. As the $J/U$ ratio decreases, while diagonal elements remain almost unchanged, off-diagonal ones decrease and finally vanish at $J/U=0$. The situation is the same for other wave vectors. Therefore, diagonal spin susceptibilities play dominant roles in determining magnetism. Fig.~\ref{fig:Pauli3D}~({\bf e}) and~({\bf f}) show two diagonal q-dependent spin susceptibilities $\chi^{33;33}_{s,RPA}(q,\omega=0)$ and $\chi^{44;44}_{s,RPA}(q,\omega=0)$ at the Fermi level. We find that pronounced peaks are preserved at the same wave vectors in respective orbitals as what is detected in orbitally resolved Pauli susceptibility. We have also checked results at different values of $U$ and found that the situation is the same. Therefore, we conclude that for reasonable strength of interactions~\cite{RPArelated} the magnetic instability in spin susceptibility remains at the same wave vector as the condensation of particle-hole excitations in Pauli susceptibility.

Next, we study the evolution of particle-hole excitations in Pauli susceptibility as a function of shifted Fermi energy.
Fig.~\ref{fig:ShiftEf} shows momentum dependent Pauli susceptibility of d$_{xz}$ and d$_{x^2-y^2}$ orbitals along the path of $(0,0,\pi)\rightarrow(\pi,0,\pi)\rightarrow(\pi,\pi,\pi)\rightarrow(0,0,\pi)$ at different Fermi energy shifts, corresponding to
different numbers of excess Fe. Without shift or as the shifts are small, like $\Delta E_F=0$ and $0.045$, both $\chi^{33;33}_{0}$ and $\chi^{44;44}_{0}$ show dominant peaks around $(\pi,\pi,\pi)$. While the peak in $\chi^{44;44}_{0}$ is slight enhanced at $\Delta E_F=0.045$, compared to the case of $\Delta E_F=0$, that in $\chi^{33;33}_{0}$ is strongly suppressed. At $\Delta E_F=0.065,0.075,0.085$, a well-defined peak centered at $(\pi,0,\pi)$ appears in $\chi^{33;33}_{0}$, indicating that the BAF state has its itinerant origin. The peak at $(\pi,\pi,\pi)$ in $\chi^{44;44}_{0}$, however, remains almost unchanged at these $\Delta E_F$.
Further increasing $\Delta E_F$ to $0.105,0.12,0.15$, the peak around $(\pi,\pi,\pi)$ in $\chi^{44;44}_{0}$ rapidly decreases, while
the peak previously right at $(\pi,0,\pi)$ in $\chi^{33;33}_{0}$ moves to an incommensurate wave vector, which is consistent with the experimental
finding of transitions from BAF states to incommensurate phases as the number of excess Fe increases~\cite{Bao2009,Li2009,Rodriguez2011,Roessler2011}. The larger the shift is, the farther away the peak is from $(\pi,0,\pi)$, which is also consistent with experiments. The incommensurate wave vector at $(\pi,\epsilon,\pi)$ indicates that magnetic moments in each row along the $a$ axis are antiparallel to each other, which modulates with propagating vector $2\pi\epsilon/b$ along the $b$ axis. Here we pointed out that experimentally observed $(\delta,0,\pi)$ incommensurability may be related to the peak at incommensurate wave vector in $\chi^{44;44}_{0}$. We also check the temperature effect which can be effectively viewed as lift of the Fermi level. As expected, the peak moves away from $(\pi,0,\pi)$ as temperature increases, which also agrees with the experimental results~\cite{Parshall2012}.

\begin{table}[!h]
\caption{Comparison of energies among different antiferromagnetic states. The collinear antiferromagnetic (CAF), plaquette antiferromagnetic (PAF), bicollinear antiferromagnetic (BAF) and N\'{e}el ordered antiferromagnetic (NAF) states are taken into account within a supercell. Both local spin density approximation (LDA) and spin polarized generalized gradient approximation (GGA) are used. FeTe denotes a supercell of Fe$_{16}$Te$_{16}$. Fe$_{1.0625}$Te means Fe$_{17}$Te$_{16}$ while Zn$_{0.0625}$FeTe indicates ZnFe$_{16}$Te$_{16}$. (see appendix \ref{sec:six})}
\begin{tabular}{cccccccccc}
\hline\hline
& $E_{CAF}-E_{NAF}$ & $E_{PAF}-E_{NAF}$ & $E_{BAF}-E_{NAF}$ \\
\hline
FeTe(LDA) & \textbf{-136.01} & -111.84 & -121.25 \\
FeTe(GGA) & -138.22 & -144.51 & \textbf{-162.83} \\
\hline
Fe$_{1.0625}$Te \\ (LDA) & -126.41 & -114.24 & \textbf{-135.81} \\
Fe$_{1.0625}$Te \\ (GGA) & -130.13 & -151.00 & \textbf{-175.84} \\
\hline
Zn$_{0.0625}$FeTe \\ (LDA) & \textbf{-128.79} & -96.91 & -107.19 \\
\hline
\end{tabular}\label{Ediff}
\end{table}

In spite of good agreements with various experiments, our itinerant picture of magnetism in Fe$_{1+x}$Te still encounters a severe problem. That is why $(\pi,0,\pi)$ instability wins the competition with $(\pi,\pi,\pi)$ instability which is obviously stronger than its rival. The reason is that excess Fe not only contributes extra electrons to the Fe plane which induces $(\pi,0,\pi)$ instability, but also provides a magnetic ion strongly coupled with in-plane Fe which stabilizes the BAF state, rather than the CAF one~\cite{Zhang2009}. In table 1, we show the importance of the interstitial magnetic ion. We construct a supercell with 16 in-plane Fe. With such a supercell, various magnetic orders like N\'{e}el antiferromagnetic (NAF), collinear antiferromagnetic (CAF), BAF, and plaquette antiferromagnetic (PAF) order, can be studied on the same foot. (see appendix \ref{sec:six} for the cartoons) Without excess Fe, the magnetic ground state is strongly dependent on the functional. Local density approximation (LDA) yields a CAF state while generalized gradient approximation (GGA) favors a BAF state. This inconsistency casts doubt on the reliability of conclusions from previous DFT investigations on FeTe where only GGA is used~\cite{Ma2009,Moon2010}. After putting one excess Fe into interstitial with the position according to neutron diffraction experiments~\cite{Li2009}, we find that the BAF state becomes the ground state irrespective of which approximation one chooses, indicating robustness of the BAF state after involving excess Fe. The crucial role of existing magnetic ion in the interstitial can be verified by changing the interstitial Fe to Zn which contributes extra electrons with nonmagnetic ions. We find that the magnetic ground state becomes CAF within LDA. This is clear evidence that the existence of an interstitial magnetic ion is crucial for stabilizing the BAF state. On the other hand, $(\pi,0,\pi)$ instability also indicates a tendency towards the PAF state in principle, in addition to the BAF state. However, our results do not support the existence of the PAF state.

Finally, we investigate the evolution of particle-hole excitations in Pauli susceptibility as a function of Te height measured from the Fe plane. We fix the number of extra electrons to be 0.2e/Fe and tune the Te height from $h_{Te}=1.8\AA$ to $1.52\AA$ in the interval of $0.04\AA$. From Fig.~\ref{fig:ShifthTe}~({\bf a}), we find that lowering Te height rapidly suppresses the $(\pi,0,\pi)$ instability in the d$_{xz}$ orbital. The peak position begins to move away from $(\pi,0,\pi)$ towards $(\delta,0,\pi)$ at $h_{Te}=1.72\AA$. The $\chi^{33;33}_{0}$ becomes featureless with further decreasing Te height, suggesting vanish of magnetic instability, which is consistent with the experiments where magnetic order disappears with substitution of Te by Se~\cite{Martinelli2010,Medvedev2010}. Note that since Se height is much lower than Te height, the instability to the BAF state is suppressed. Meanwhile, reduction of Te height also gradually suppresses the $(\pi,\pi,\pi)$ instability in the d$_{x^2-y^2}$ orbital as shown in Fig.~\ref{fig:ShifthTe}~({\bf b}). However, it remains relatively large either at or close to $(\pi,\pi,\pi)$. Therefore, as the $(\pi,0,\pi)$ instability of $\chi^{33;33}_{0}$ is completely suppressed, the remaining $(\pi,\pi,\pi)$ instability of $\chi^{44;44}_{0}$ may be the source of superconductivity observed experimentally~\cite{Liu2010}, since its pairing is also associated with the real part of Pauli susceptibility through an integral over the Fermi surface.

\section{Conclusions}
In summary, our results offer a possible explanation of why the Fermi surface is nested at $(\pi,\pi)$ while magnetic order at $(\pi,0)$ in Fe$_{1+x}$Te, as a hidden $(\pi,0)$ instability coexists with the nesting of the Fermi surface at $(\pi,\pi)$ within a reasonable range of extra electrons donated from excess Fe and height of Te ion measured from the Fe plane. Existence of interstitial magnetic Fe further selects the $(\pi,0)$ instability as the itinerant origin of the BAF state. Increasing excess Fe and decreasing the height of the chalcogen atom both suppress the $(\pi,0)$ instability and may favor, respectively, the incommensurate AF states and the superconducting states with pairing mediated by $(\pi,\pi)$ magnetic fluctuations. Even though there exists a diversity of magnetic orders in iron pnictides and iron chalcogenides, their origins can be explained from both itinerant and localized limit, suggesting that a unified model for describing iron-based superconductors should involve both itinerant electrons and local spins. On the other hand, orbital differentiation has to be seriously taken into account, since different orbitals play different roles in magnetism and superconductivity as seen in Fe$_{1+x}$Te where evolutions of Pauli susceptibilities of d$_{xz}$/d$_{yz}$ and d$_{x^2-y^2}$ are remarkably different when filling or structure is changed. Our findings may also point out a way of understanding magnetism and superconductivity in other iron-based superconductors, besides the existing localized scenario~\cite{Hu2012}.

\section{Acknowledgments}
YZ is supported by National Natural Science Foundation of China (No. 11174219), Shanghai Pujiang Program (No. 11PJ1409900), Research Fund for the Doctoral Program of Higher Education of China (No. 20110072110044), the Program for Professor of Special Appointment (Eastern Scholar) at Shanghai Institutions of Higher Learning, and is indebted to CSRC for the hospitality and partial financial support from CAEP.

\appendix
\section{Possible valence of interstitial iron}
\label{sec:one}
\begin{figure}
\includegraphics[width=\columnwidth]{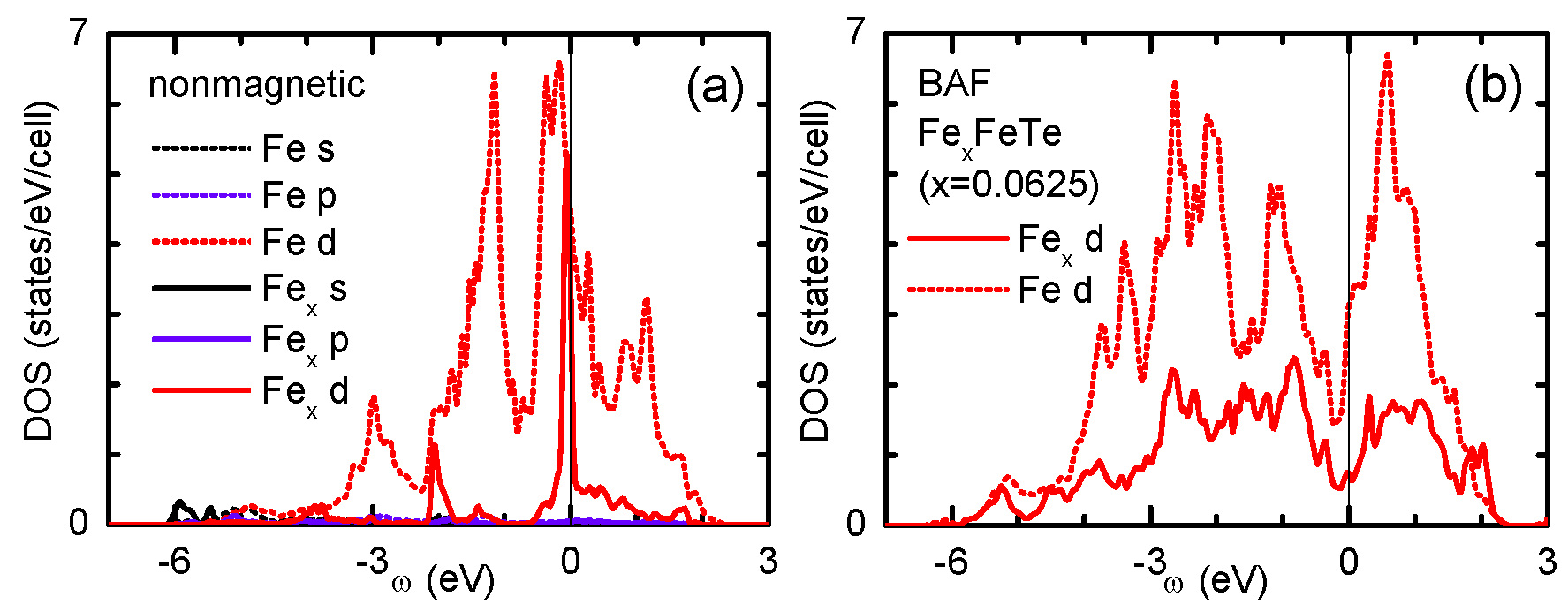}
\caption{(Color online)\textbf{Possible valence of interstitial iron}.(\textbf{a}) Partial density of state in nonmagnetic state. Solid line denotes the contributions from interstitial iron while dotted line the contributions from all the in-plane iron divided by a factor of 2. (\textbf{b}) Partial density of state in bicollinear antiferromagnetic state. Now the contributions from all the in-plane iron are divided by a factor of 4. The supercell we used is Fe$_{17}$Te$_{16}$.} \label{fig:valence}
\end{figure}

In order to estimate possible valence of interstitial Fe, we analyze partial density of states obtained from both nonmagnetic state and bicollinear antiferromagnetic state of a supercell of Fe$_{17}$Se$_{16}$ with sixteen in-plane Fe and one interstitial Fe. The supercell is constructed according to experimental structure of Fe$_{1.068}$Te at 80K~\cite{Li2009}. The results from first principles calculations~\cite{Blaha2001} are presented in Fig.~\ref{fig:valence}. We find that 3d orbitals of interstitial Fe are also prominently occupied, indicating that interstitial Fe should not be in a valence state of $+8$ where 3d orbitals of interstitial Fe should be empty. Instead, we find that the occupation number on interstitial Fe is almost the same as that on in-plane Fe by comparing the integrated density of state under Fermi level for all the irons. Therefore, possible valence of interstitial iron should be the same as that of in-plane Fe, i.e. $\sim +2$. Moreover, the density of state of interstitial Fe in nonmagnetic state (Fig.~\ref{fig:valence}~({\bf a})) shows a strong peak at Fermi level, indicating a strong tendency towards a magnetically ordered state.

\section{Shift of Fermi level VS number of excess iron}
\label{sec:two}

\begin{figure}
\includegraphics[width=\columnwidth]{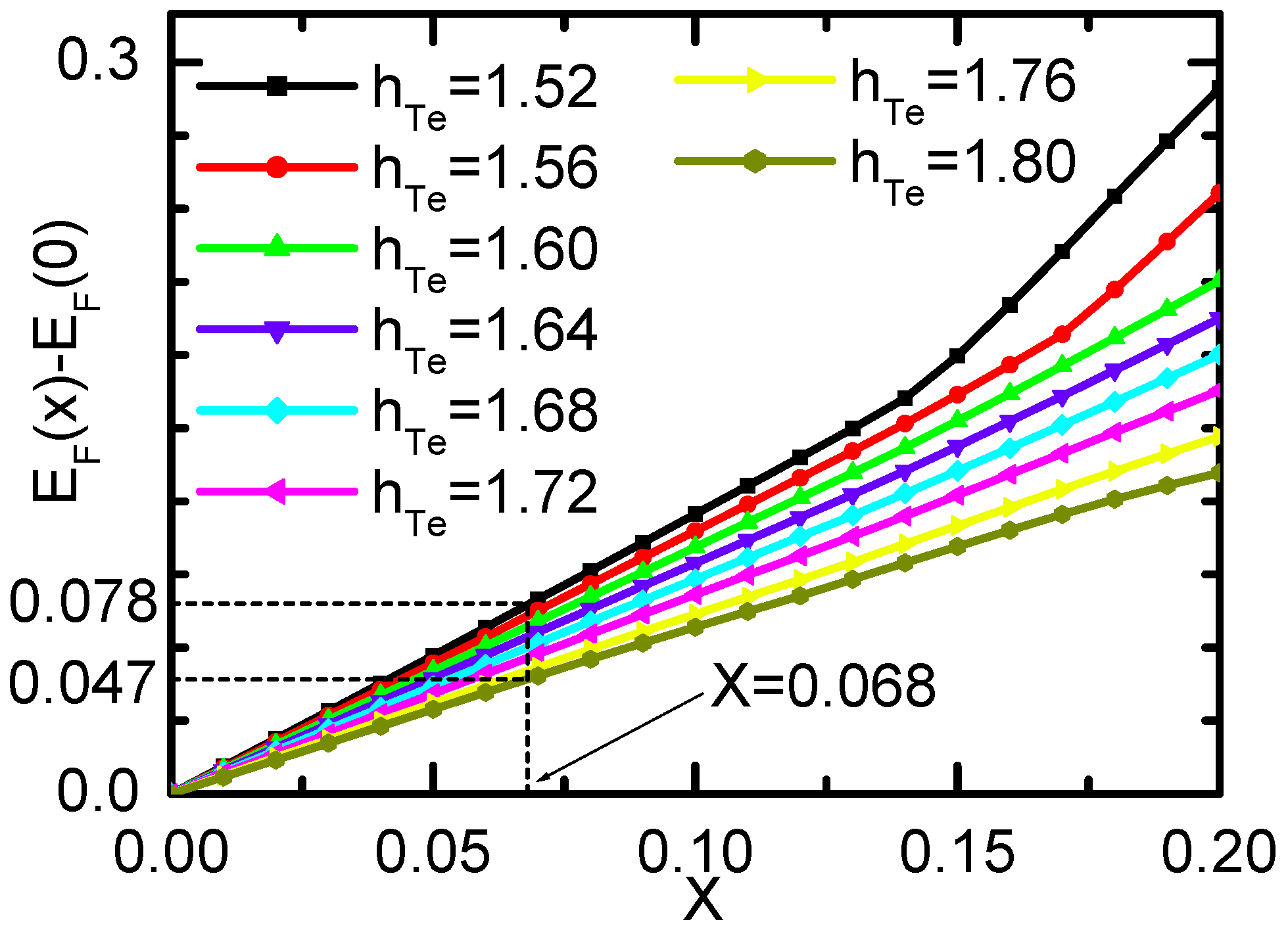}
\caption{(Color online)\textbf{Shift of Fermi level as a function of excess iron concentration}. Experimental lattice constants are used while height of Te ion measured from Fe plane is varied from 1.52 $\AA$ to 1.80 $\AA$ in a interval of 0.04 $\AA$. The Shift of Fermi level of Fe$_{1+x}$Te compound at $x=0.068$ calculated from density functional theory is one order of magnitude smaller than what has been done by Han and Savrasov~\cite{Han2009,Singh2010,Han2010}.} \label{fig:ShiftEfVSx}
\end{figure}

Based on the fact that valence of interstitial Fe is $+2$, extra electrons doped into Fe plane should be counted as $2x$ in Fe$_{1+x}$Te compounds, rather than $8x$~\cite{Han2009,Singh2010,Han2010}. By rigid band shifts, in Fig.~\ref{fig:ShiftEfVSx}, we show shift of Fermi level as a function of concentration of excess irons. The lattice constants are taken from experiments~\cite{Li2009}. The height of Te ion measured from Fe plane is varied from 1.52 $\AA$ to 1.80 $\AA$ in a interval of 0.04 $\AA$, since it is well known that the electronic properties of Fe-based superconductors are strongly dominated by the height of anion. By applying first principles calculations, we find from Fig.~\ref{fig:ShiftEfVSx} that the shift of Fermi level should be one order of magnitude smaller than what has been done by Han and Savrasov~\cite{Han2009,Singh2010,Han2010}.

\section{Fermi surface}
\label{sec:three}
\begin{figure}
\includegraphics[width=\columnwidth]{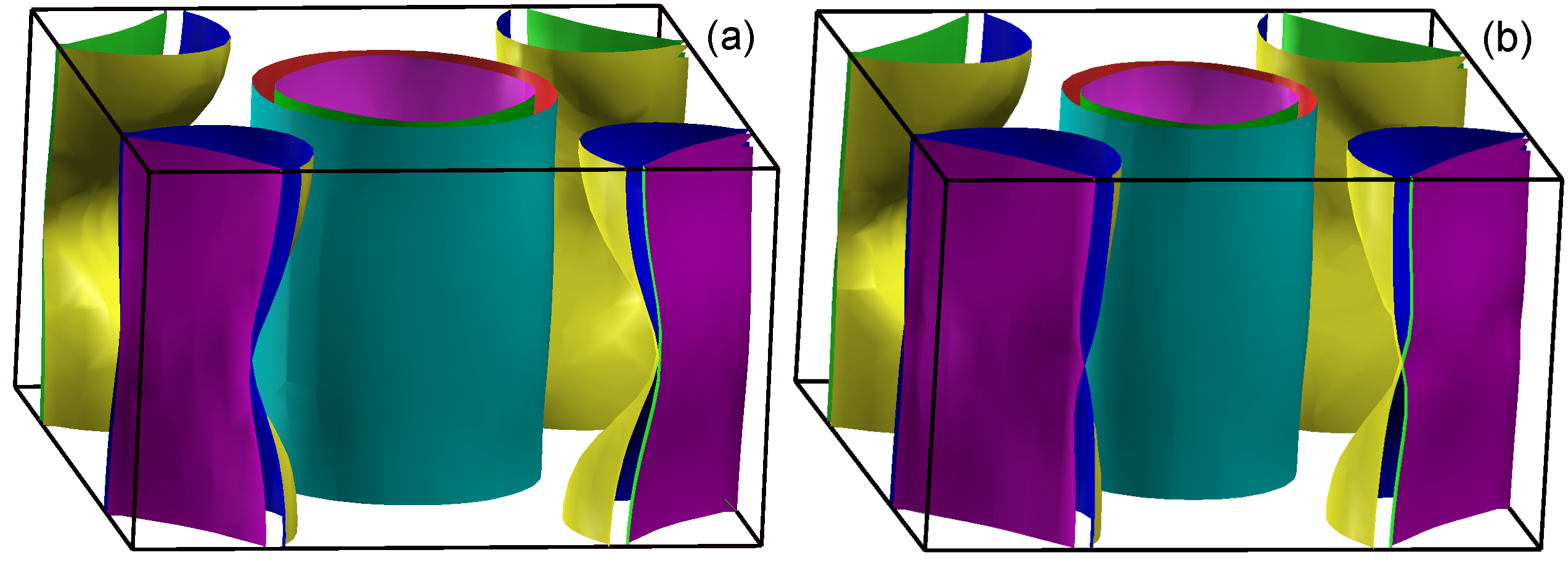}
\caption{(Color online)\textbf{Fermi surfaces}.(\textbf{a}) Without shift of Fermi level. (\textbf{b}) Fermi level is shifted up by 0.075eV. Besides the difference in the radius of Fermi cylinders at the $\Gamma$ and $M$ points, the overall shapes of Fermi surfaces in these two cases are quite similar. Experimental lattice structure at high temperature is used~\cite{Li2009}. } \label{fig:Fermisurfaces}
\end{figure}

In Fig.~\ref{fig:Fermisurfaces}, we compare the Fermi surfaces in the absence and presence of Fermi energy shift. We find that with a reasonable Fermi energy shift, such as 0.075eV, while the radius of Fermi cylinders at the $\Gamma$ and $M$ point are different, the overall shapes of Fermi surfaces are still the same, indicating that Fermi surface topology is still consistent with that observed by angular resolved photoemission spectroscopy even though there is a shift of Fermi energy due to the extra electrons contributed from interstitial excess Fe.

\section{Pauli susceptibility within Constant matrix element approximation}
\label{sec:four}
\begin{figure}
\includegraphics[width=\columnwidth]{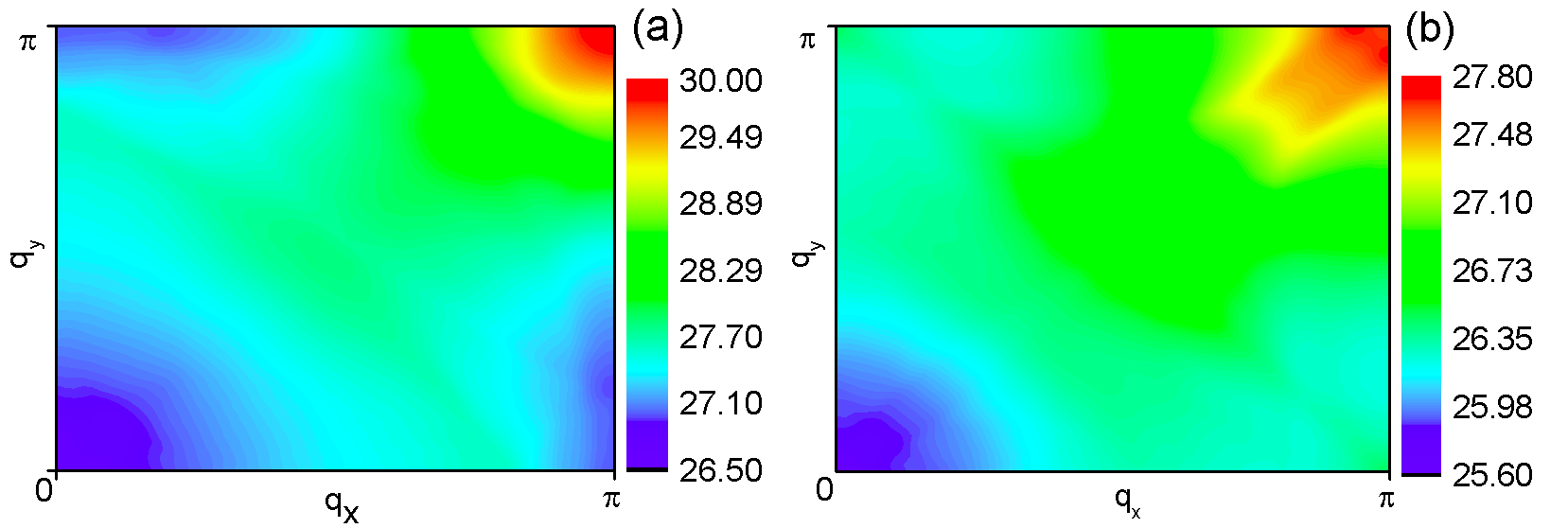}
\caption{(Color online)\textbf{Pauli susceptibilities}. Here we adopt constant matrix element approximation. (\textbf{a}) Without shift of Fermi level. (\textbf{b}) Fermi level is shifted up by 0.075eV. Experimental lattice structure at high temperature is used~\cite{Li2009}.} \label{fig:Pauli}
\end{figure}

In Fig.~\ref{fig:Pauli}, we show the results of Pauli susceptibilities within constant matrix element approximation. Without a Fermi level shift, a strong peak at $q=(\pi,\pi,\pi)$ is observed. With a Fermi level shift of 0.075eV, the peak at $q=(\pi,\pi,\pi)$ becomes a bit broader. However£¬no prominent peak can be observed at $q=(\pi,0,\pi)$ or $q=(0,\pi,\pi)$.

\section{exact expression of Pauli susceptibility and spin susceptibility within RPA}
\label{sec:five}
The Pauli susceptibility~\cite{Graser2009} is defined as
\begin{eqnarray}
  \chi^{pq;st}_{0} &=& -\frac{1}{N} \sum_{k,\mu\nu}\frac{a^{s}_{\mu}(k)a^{p*}_{\mu}(k)a^{q}_{\nu}(k+q)a^{t*}_{\nu}(k+q)}{\omega+E_{\nu}(k+q)-E_{\mu}(k)+i0^{+}} \nonumber \\
  &&\times [f(E_{\nu}(k+q))-f(E_{\mu}(k))]
\end{eqnarray}
where matrix elements $a^{s}_{\mu}(k)=\langle s|\mu k \rangle$ connect the orbital and the band spaces and are the components of the eigenvectors resulting from the diagonalization of the effective tight-binding Hamiltonian. Here $f(E)$ is the Fermi distribution function. The spin susceptibility~\cite{Graser2009} within random phase approximation (RPA) is obtained in the form of Dyson type equations as
\begin{eqnarray}
  \chi^{pq;st}_{s,RPA} &=& \chi^{pq;st}_{0} + \chi^{pq;uv}_{s,RPA} U_s^{uv;wz} \chi^{wz;st}_{0}
\end{eqnarray}
where repeated indices are summed over.

\section{Orbital resolved Pauli susceptibility and spin susceptibility}
\label{sec:fourx}
\begin{figure}
\includegraphics[width=\columnwidth]{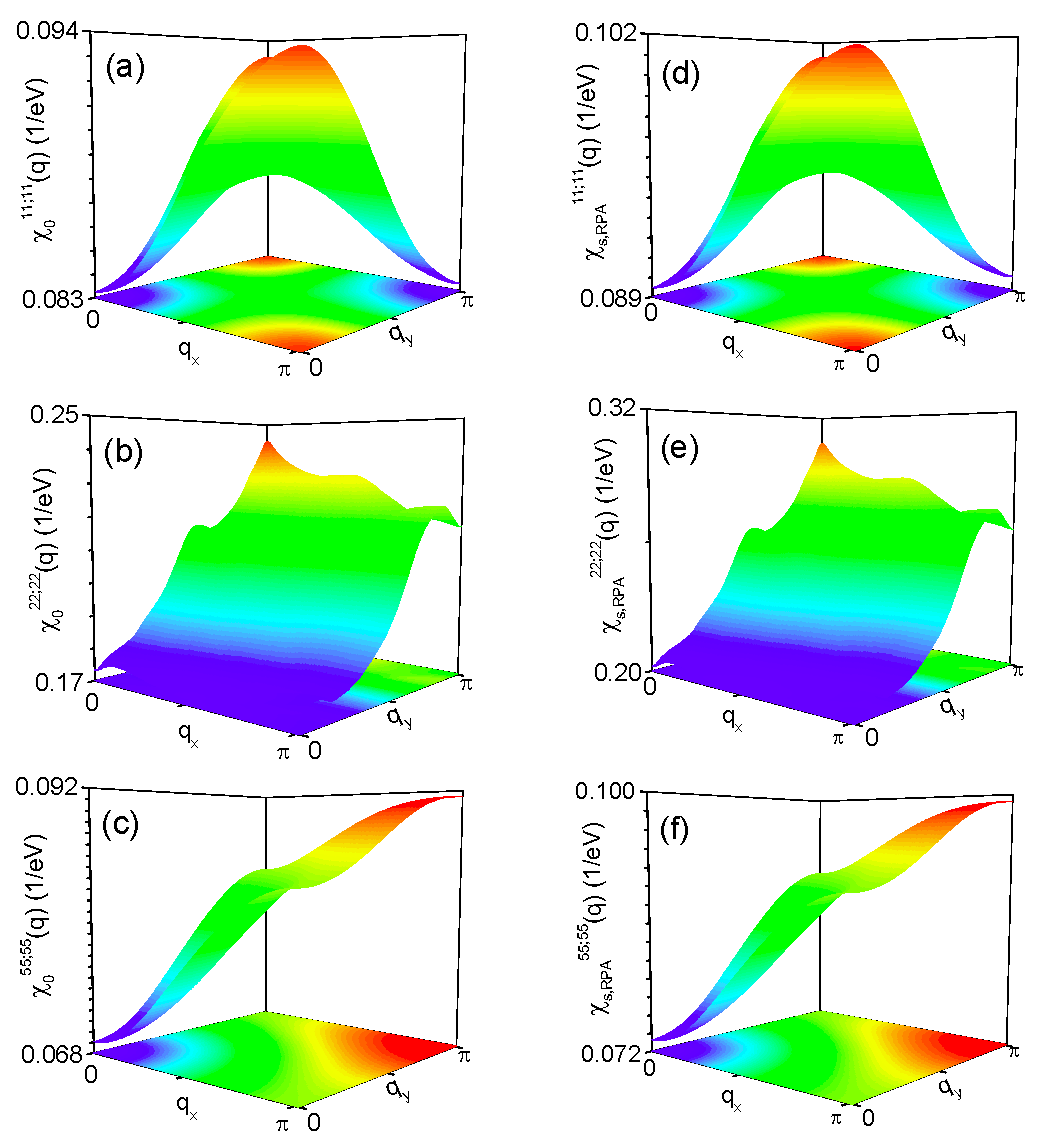}
\caption{(Color online){\bf Orbitally resolved Pauli susceptibility and spin susceptibility}. q-dependent Pauli susceptibilities $\chi^{11;11}_{0}(q,\omega=0)$ (\textbf{a}), $\chi^{22;22}_{0}(q,\omega=0)$ (\textbf{b}) and $\chi^{55;55}_{0}(q,\omega=0)$ (\textbf{c}). q-dependent spin susceptibilities $\chi^{11;11}_{s,RPA}(q,\omega=0)$ (\textbf{d}), $\chi^{22;22}_{s,RPA}(q,\omega=0)$ (\textbf{e}) and $\chi^{55;55}_{s,RPA}(q,\omega=0)$ (\textbf{f}). Here Fermi level is shifted up by 0.075eV. Spin susceptibility is calculated at $J=U/4$ and $U=0.8$ eV.} \label{fig:PauliandSpin}
\end{figure}

In Figs.~\ref{fig:PauliandSpin} (a)-(c), three elements of Pauli susceptibility $\chi^{tt;tt}_{0}(q,\omega=0)$ with $t=1,2,5$ which is relevant to magnetism are shown. In Figs.~\ref{fig:PauliandSpin} (d)-(f), corresponding spin susceptibilities are present.

\section{Cartoons for different magnetic states}
\label{sec:six}
\begin{figure}
\includegraphics[width=0.7\columnwidth]{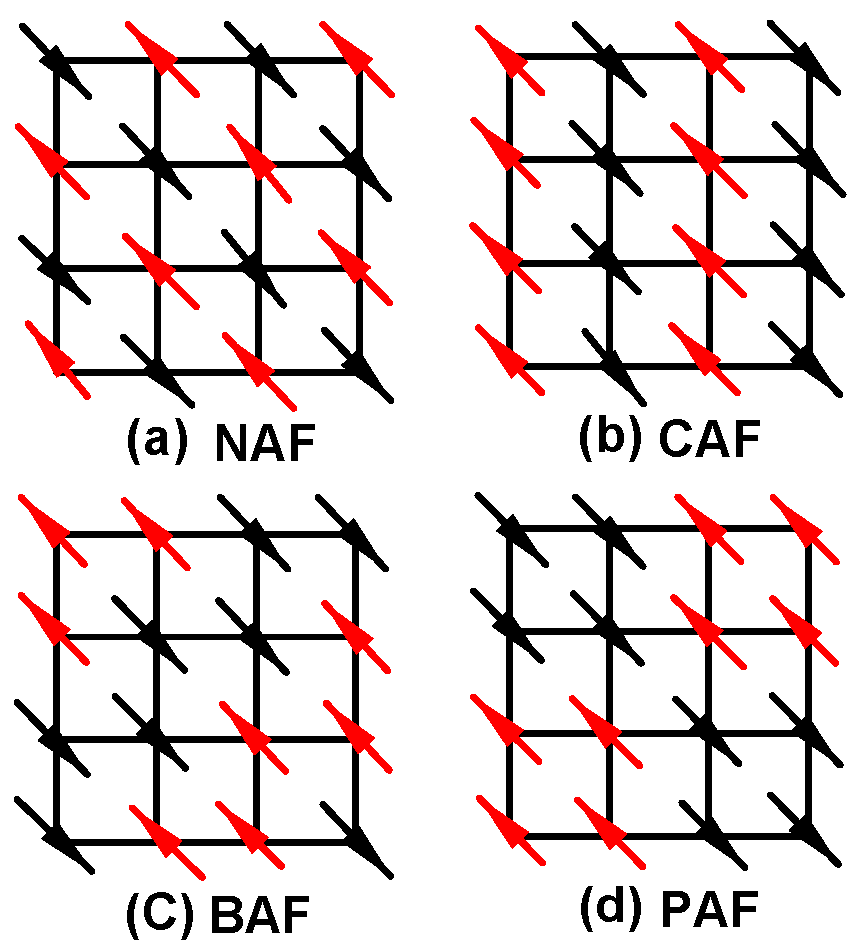}
\caption{(Color online)\textbf{Cartoons for different magnetic states}.(\textbf{a}) Neel ordered antiferromagnetic (NAF) state. (\textbf{b}) Collinear antiferromagnetic (CAF) state. (\textbf{a}) Bicollinear antiferromagnetic (BAF) state. (\textbf{b}) Plaquette antiferromagnetic (PAF) state.} \label{fig:Cartoons}
\end{figure}

In Fig.~\ref{fig:Cartoons}, we show the cartoons for different magnetic states we considered in this investigation. Please note, the $(\pi,0)$ instability favors two kinds of magnetically ordered states. One is the bicollinear antiferromagnetic (BAF) state (Fig.~\ref{fig:Cartoons}~({\bf c})), the other is the plaquette antiferromagnetic (PAF) state (Fig.~\ref{fig:Cartoons}~({\bf d})). From our first principles calculations, the energy of the BAF state is always lower than that of the PAF state in all the cases.

\section{comparison of Band structure and DOS from LDA and GGA}
\label{sec:seven}
\begin{figure}
\includegraphics[width=\columnwidth]{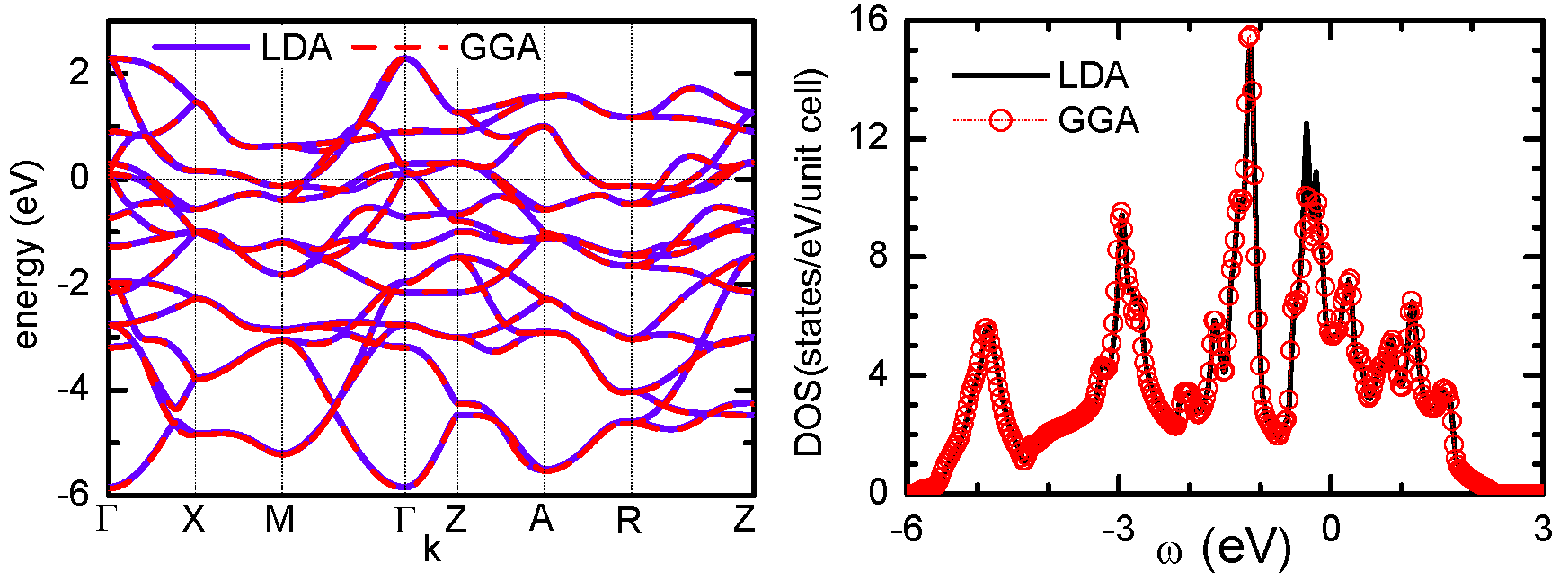}
\caption{(Color online)\textbf{Band structure and DOS from LDA and GGA}.(\textbf{a}) Comparison of band structure from local density approximation (LDA) and generalized gradient approximation (GGA). (\textbf{b}) Comparison of DOS from LDA and GGA.} \label{fig:LDAvsGGA}
\end{figure}

Fig.~\ref{fig:LDAvsGGA} shows the comparisons of band structure, as well as density of state, obtained from local density approximation (LDA) and generalized gradient approximation (GGA). The results from these two approximation are perfectly consistent with each other, indicating that our results are independent of which approximation we choose.

\begin{figure}
\includegraphics[width=\columnwidth]{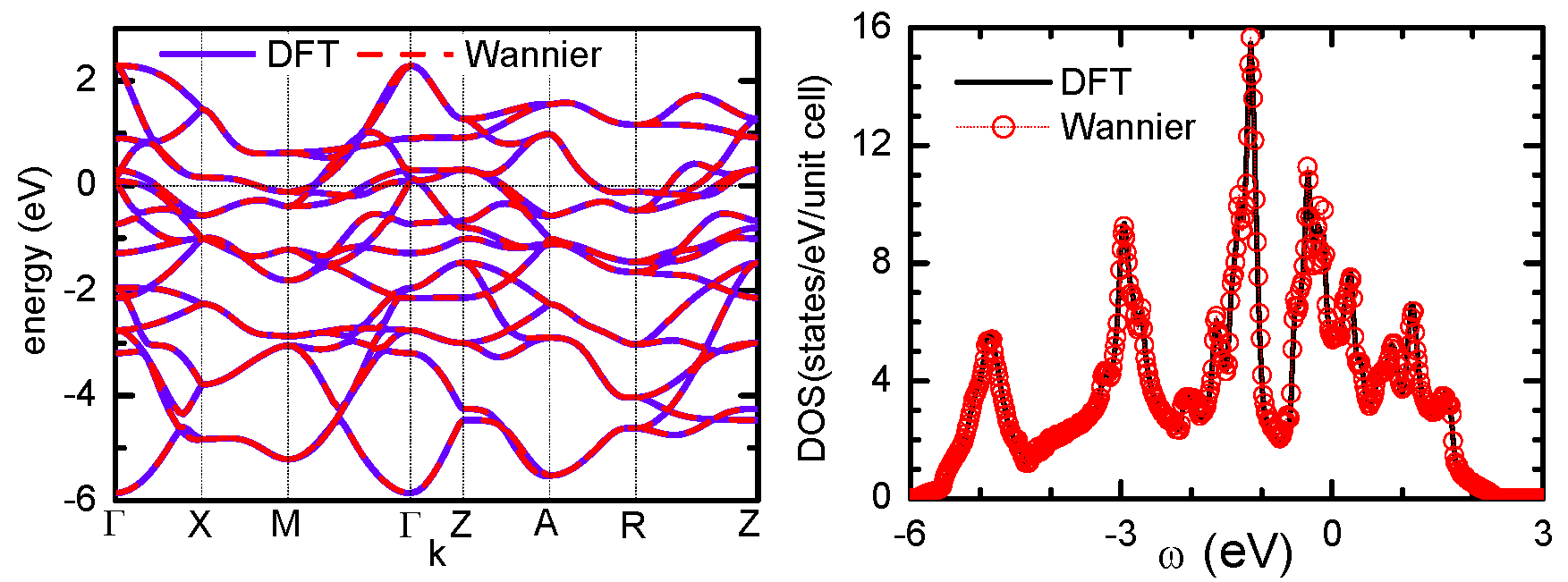}
\caption{(Color online)\textbf{Band structure and DOS from DFT and effective model}.(\textbf{a}) Comparison of band structure calculated from density functional theory and effective tight banding model. (\textbf{b}) Comparison of DOS calculated from density functional theory and effective tight banding model.} \label{fig:DFTvswannier}
\end{figure}

\section{comparison of Band structure and DOS from DFT and effective model}
\label{sec:eight}
Fig.~\ref{fig:DFTvswannier} presents the comparisons of band structure, as well as density of state, obtained from first principles calculations and tight-binding model calculations. The effective tight-binding model is derived from first principles band structure through a transformation from Bloch space to maximally localized Wannier orbital basis. The energy window we used is from -6 eV to 3 eV where the weight of each eigenstate is mostly contributed from Fe 3d orbitals and Te 4p orbitals. In order to get a perfect consistency in band structure calculated from first principles and the effective tight-binding model, we keep the long-range hoppings from the atoms in the unit cell at $[0,0,0]$ to those at $[9\mathbf{a},9\mathbf{b},5\mathbf{c}]$. The results calculated from density functional theory and from the effective tight-binding model are perfectly consistent with each other, indicating that the transformation we used will not impose any systematical error on our results.


\begin{thebibliography}{99}

\bibitem{Liu2010} T. J. Liu, J. Hu, B. Qian, D. Fobes, Z. Q. Mao1,W. Bao, M. Reehuis, S. A. J. Kimber, K. Proke, S. Matas, D. N. Argyriou, A. Hiess, A. Rotaru, H. Pham, L. Spinu, Y. Qiu, V. Thampy, A. T. Savici, J. A. Rodriguez and C. Broholm, Nature Materials {\bf 9}, 716 (2010).

\bibitem{Zhao2008} J. Zhao, Q. Huang, C. De La Cruz, S. Li, J. W. Lynn, Y. Chen, M. A. Green, G. F. Chen, G. Li, Z. Li, J. L. Luo, N. L. Wang and P. Dai, Nature Materials {\bf 7}, 953 (2008).

\bibitem{Kimber2009} S. A. J. Kimber, A. Kreyssig, Y.-Z. Zhang, H. O. Jeschke, R. Valent\'\i, F. Yokaichiya, E. Colombier, J. Yan, T. C. Hansen, T. Chatterji, R. J. McQueeney, P. C. Canfield, A. I. Goldman and D. N. Argyriou, Nature Materials {\bf 8}, 471 (2009).

\bibitem{Paglione2010} J. Paglione, R. L. Greene, Nature Physics {\bf 6}, 645 (2010).

\bibitem{Mazin2010} I. I. Mazin, Nature (London) {\bf 464}, 183 (2010).

\bibitem{Mazin2008} I. I. Mazin, D. J. Singh, M. D. Johannes, M. H. Du, Phys. Rev. Lett. {\bf 101}, 057003 (2008).

\bibitem{Yildirim2008} T. Yildirim, Phys. Rev. Lett. {\bf 101}, 057010 (2008).

\bibitem{Si2008} Q. Si, E. Abrahams, Phys. Rev. Lett. {\bf 101}, 076401 (2008).

\bibitem{Ma2009} F. J. Ma, W. Ji, J. Hu, Z.-Y. Lu, T. Xiang, Phys. Rev. Lett. {\bf 102}, 177003 (2009).

\bibitem{Moon2010} C. Y. Moon, H. J. Choi, Phys. Rev. Lett. {\bf 104}, 057003 (2010).

\bibitem{Xia2009} Y. Xia, D. Qian, L. Wray, D. Hsieh, G. F. Chen, J. L. Luo, N. L. Wang, and M. Z. Hasan, Phys. Rev. Lett. {\bf 103}, 037002 (2009).

\bibitem{Nakayama2010} K. Nakayama, T. Sato, P. Richard, T. Kawahara, Y. Sekiba, T. Qian, G. F. Chen, J. L. Luo, N. L. Wang, H. Ding, and T. Takahashi, Phys. Rev. Lett. {\bf 105}, 197001 (2010).

\bibitem{Bao2009} W. Bao, Y. Qiu, Q. Huang, M. A. Green, P. Zajdel, M. R. Fitzsimmons, M. Zhernenkov, S. Chang, M. Fang, B. Qian, E. K. Vehstedt, J. Yang, H. M. Pham, L. Spinu, and Z. Q. Mao, Phys. Rev. Lett. {\bf 102}, 247001 (2009).

\bibitem{Li2009} S. Li, C. de la Cruz, Q. Huang, Y. Chen, J. W. Lynn, J. Hu, Y.-L. Huang, F.-C. Hsu, K.-W. Yeh, M.-K. Wu, and P. Dai, Phys. Rev. B {\bf 79}, 054503 (2009).

\bibitem{Kamihara2008} Y. Kamihara, T. Watanabe, M. Hirano, H. Hosono, J. Am. Chem. Soc. {\bf 130}, 3296 (2008).

\bibitem{Rotter2008} M. Rotter, M. Tegel, D. Johrendt, Phys. Rev. Lett. {\bf 101}, 107006 (2008).

\bibitem{Hsu2008} F.-C. Hsu, J.-Y. Luo, K.-We. Yeh, T.-K. Chen, T.-W. Huang, P. M. Wu, Y.-C. Lee, Y.-L. Huang, Y.-Y. Chu, D.-C. Yan and M.-K. Wu, Proc. Natl Acad. Sci. USA {\bf 105}, 14262 (2008).

\bibitem{Han2009} M. J. Han, S. Y. Savrasov, Phys. Rev. Lett. {\bf 103}, 067001 (2009).

\bibitem{Singh2010} P. P. Singh, Phys. Rev. Lett. {\bf 104}, 099701 (2010).

\bibitem{Han2010} M. J. Han, S. Y. Savrasov, Phys. Rev. Lett. {\bf 104}, 099702 (2010).

\bibitem{Rodriguez2011} E. E. Rodriguez, C. Stock, P. Zajdel, K. L. Krycka, C. F. Majkrzak, P. Zavalij, and M. A. Green, Phys. Rev. B {\bf 84}, 064403 (2011).

\bibitem{Dong2008} J. Dong, H. J. Zhang, G. Xu, Z. Li, G. Li, W. Z. Hu, D. Wu, G. F. Chen, X. Dai, J. L. Luo, Z. Fang and N. L. Wang, EPL {\bf 83}, 27006 (2008).

\bibitem{Heil2012} In arXiv:1210.2593, matrix elements are involved in the calculation of Pauli susceptibility without orbital resolution.

\bibitem{Blaha2001} P. Blaha, K. Schwarz, G. Madsen, D. Kvaniscka, and J. Luitz, WIEN2K, {\it An Augmented Plane Wave+Local Orbitals Program for Calculating Crystal}, edited by K. Schwarz (Techn. University, Vienna, Austria, 2001).

\bibitem{Wannier90} A. A. Mostofi, J. R. Yates, Y.-S. Lee, I. Souza, D. Vanderbilt and N. Marzari, Comput. Phys. Commun. {\bf 178}, 685 (2008).

\bibitem{Wientowannier} J. Kunes, R. Arita, P. Wissgott, A. Toschi, H. Ikeda, K. Held, Comput. Phys. Commun. {\bf 181}, 1888 (2010).

\bibitem{Graser2009} S. Graser, T. A. Maier, P. J. Hirschfeld and D. J. Scalapino, New Journal of Physics {\bf 11}, 025016 (2009).

\bibitem{Figureref} The concentration $x$ of excess Fe as a function of shifted Fermi level is calculated, based on the experimental structure of Fe$_{1.068}$Te at $80$ K as given in Ref.~\onlinecite{Li2009}, within local density approximation where dynamical correlations are completely ignored~\cite{Han2010}.

\bibitem{RPArelated} In the RPA where the self-energy correction in the Green's function is neglected, we have to take small values of $U$ to ensure all the elements of spin susceptibility are away from their critical regions~\cite{Kuroki2008}.

\bibitem{Kuroki2008} K. Kuroki, S. Onari, R. Arita, H. Usui, Y. Tanaka, H. Kontani, and H. Aoki, Phys. Rev. Lett. {\bf 101}, 087004 (2008).

\bibitem{Roessler2011} S. R\"{o}{\ss}ler, D. Cherian, W. Lorenz, M. Doerr, C. Koz, C. Curfs, Yu. Prots, U. K. R\"{o}{\ss}ler, U. Schwarz, S. Elizabeth, and S. Wirth, Phys. Rev. B {\bf 84}, 174506 (2011).

\bibitem{Parshall2012} D. Parshall, G. Chen, L. Pintschovius, D. Lamago, Th. Wolf, L. Radzihovsky, and D. Reznik, Phys. Rev. B {\bf 85}, 140515(R) (2012).

\bibitem{Zhang2009} L. Zhang, D. J. Singh, M. H. Du, Phys. Rev. B {\bf 79}, 012506 (2009).

\bibitem{Martinelli2010} A. Martinelli, A. Palenzona, M. Tropeano, C. Ferdeghini, M. Putti, M. R. Cimberle, T. D. Nguyen, M. Affronte, and C. Ritter, Phys. Rev. B {\bf 81}, 094115(R) (2010).

\bibitem{Medvedev2010} S. Medvedev, T. M. McQueen, I. A. Troyan, T. Palasyuk, M. I. Eremets, R. J. Cava, S. Naghavi, F. Casper, V. Ksenofontov, G. Wortmann and C. Felser, Nature Materials {\bf 8}, 630-633 (2010).

\bibitem{Hu2012} J. Hu and H. Ding, Sci. Rep. {\bf 2} 381, (2012).

\end{thebibliography}
\end{document}